\documentstyle[prl,aps,preprint, epsf]{revtex}
\begin{document}

\title{On stimulated transitions between the self-trapped states
of the nonlinear Schr\"odinger equation}

\author{P. V. Elyutin \footnote {E-mail address: pve@astra.phys.msu.su}
and A.N. Rogovenko}
\address{Department of Physics, Moscow State University, Moscow
119899, Russia}

\maketitle

\begin{abstract}
The studied model describes a particle that obeys a one-dimensional nonlinear
Schr\"odinger equation in the potential of a double-well.  Transitions between
the two lowest self-trapped states of this system under the influence of the external
time-dependent perturbation are studied in the two-mode approximation.
If the perturbation dependence on time is harmonic with the frequency $\omega$,
then transitions between the states become possible if the amplitude of the
perturbation $F$ exceeds some threshold value $F_c(\omega)$; above the threshold
motion of the system becomes chaotic.  If the perturbation is a broadband noise,
then transitions between the states are possible at arbitrarily small $F$ and occur
in the process of the system's energy diffusion.
\end{abstract}

\vspace{5mm}
\noindent
PACS number(s): 03.65, 05.45

\newpage

\section{Introduction}

The nonlinear Schr\"odinger equation (NLS),
\begin{equation} \label{01}
i\hbar {{\partial \psi } \over {\partial t}}=-{{\hbar ^2} \over {2m}}
\Delta \psi +U\left( {\vec r} \right)\psi +\lambda \left| \psi \right|^2\psi,
\end{equation}
serves as a tool for the account of many physical phenomena.  The stationary version of
this equation was introduced by Deigen and Pekar to describe the self-trapped
(autolocalized) states of electrons in deformable crystal lattice (\cite {DP51}, see also
\cite {K85}).  Gross \cite {G61} and Pitayevskii \cite {P61} derived Eq. (1) as the
mean-field approximation for the macroscopic wavefunction $\psi \left( {\vec r,t}
\right)$ of the Bose - Einstein condensate of the non-ideal Bose gas at vanishing
temperature (see also \cite {DG+99}).  The third avatar of the NLS came in the realm
of the nonlinear optics, where $\psi \left( {\vec r,t} \right)$ represents the envelope
of a quasimonochromatic electromagnetic wave.  The exact solution of the
one-dimensional homogeneous ($U\left( {\vec r} \right)=0$) NLS found by Zakharov
and Shabat \cite {ZSh71} formed the modern paradigm of the soliton theory \cite {AC91}.

Although the solutions of the NLS were extensively studied, it seems that comparatively
little is known about their properties in case of time-dependent potentials.  In this paper
we address ourselves to a specific problem of this class.  We shall consider the one-dimensional
Eq. (1) with the potential that consists of two parts: a permanent potential $U\left( x \right)$
that has the form of symmetric double well, and some time-dependent potential
$V\left( {x,t} \right)$ that will be called perturbation.  In the absence of perturbation the
properties of the stationary solutions of Eq. (1), that have the form
\begin{equation} \label{02}
\psi \left( {x,t} \right)=\Phi \left( x \right)\exp \left( {-i{E \over \hbar }t}
\right),
\end{equation}
depend essentially on the nonlinearity parameter $\lambda $.  At small $\lambda $
there is an infinite set of modes (2) that have symmetric wave functions - odd or even,
$\Phi \left( x \right)=\pm \Phi \left( {-x} \right)$.  With the increase of
$\lambda $ at some threshold value of $\lambda _c$ a pair of stationary solutions (2)
with broken symmetry, $\Phi _s \left( x \right)\ne \pm \Phi _s \left( {-x}
\right)$, comes into being.  These solutions describe the states of the particle that are
self-trapped in one of the wells of the permanent potential.   For $ \lambda <0$ the
corresponding energy $E_s$ is lower than that of any symmetric mode \cite {S97}.

Our main concern will be the following question: if the initial state of the system is one
of these self-trapped states, then how the system will evolve under the influence of the
non-stationary perturbation?  In particular, can the perturbation transfer the system
completely to the opposite self-trapped state?

There is a favorable circumstance that allows to simplify the problem.  It happens that at
moderate $\lambda >\lambda _c$ even the self-trapped modes of high asymmetry can
be accurately described by linear combination of the two lowest symmetric modes, the
even $\Phi _0\left( x \right)$ and the odd $\Phi _1\left( x \right)$ \cite
{OK+99}.  Therefore studying the problem we can restrict ourselves by analysis of
the evolution of the two-level system.  In Section II we derive the basic equations for
this model.  In Section III we study the influence of the external perturbation that
harmonically depends on time.  The evolution of the system under the influence of the
broadband noise is studied in Section IV.  Section V contains the concluding
discussion.

\section{The basic equations}

For the future use we introduce the following quantities related to the (supposedly real)
eigenfunctions $\Phi _0\left( x \right)$ and $\Phi _1\left( x \right)$:
\begin{equation} \label{03}
J_{00}=\int {\Phi _0^4dx, \qquad J_{01}=\int {\Phi _0^2\Phi _1^2dx}, \qquad
J_{11}=}\int {\Phi _1^4dx}.
\end{equation}
Let's represent the wave function of the system by the superposition of two lowest
symmetric modes,
\begin{equation} \label{04}
\psi \left( {x,t} \right)=b_0\left( t \right)\Phi _0\left( x \right)e^{-i\beta
_0t}+b_1\left( t \right)\Phi _1\left( x \right)e^{-i\beta _1t},
\end{equation}
where $\beta _i=\hbar ^{-1}\left( {E_i+\lambda J_{ii}} \right)$.  By substitution of
Eq. (4) in Eq. (1), consequent multiplication by $\Phi _i\left( x \right)$, and
integration over the coordinate $x$ we get the following system of two equations for the
complex amplitudes $b_i$:
\begin{equation} \label{05}
\begin{array} {c}
i\hbar {\displaystyle {{db_0} \over {dt}}}=-\lambda b_0\left| {b_0} \right|^2J_{00}-2\lambda
b_0\left| {b_1} \right|^2J_{01}-\lambda b_0^*b_1^2J_{01}e^{i2\left( {\beta _0-\beta
_1} \right)t}\\*[9pt] \qquad +b_0V_{00}\left( t \right)+b_1V_{01}\left( t
\right)e^{i\left( {\beta _0-\beta _1} \right)t},\\

i\hbar {\displaystyle {{db_1} \over {dt}}}=-\lambda b_1\left| {b_1} \right|^2J_{11}-2\lambda
b_1\left| {b_0} \right|^2J_{01}-\lambda b_0^2b_1^*J_{01}e^{-i2\left( {\beta _0-\beta
_1} \right)t}\\*[9pt] \qquad +b_1V_{11}\left( t \right)+b_0V_{01}\left( t \right)e^{-
i\left( {\beta _0-\beta _1} \right)t},
\end{array}
\end{equation}
where the matrix elements of the perturbation are given by the integrals
\begin{equation} \label{06}
V_{ij}\left( t \right)=\int {\Phi _i\left( x \right)\hat V\left( {x,t} \right)\Phi
_j\left( x \right)dx}.
\end{equation}
The system (5) conserves the norm of the state $\psi \left( {x,t} \right)$ (the sum of
probabilities $\left| {b_0} \right|^2+\left| {b_1} \right|^2=1$), and the common
phase of the wave function (4) is physically irrelevant.  Therefore we can describe the
evolution of the system by just two real variables.  The complex amplitudes could be cast
in the form $b_i=\sqrt {n_i}\exp (-i\vartheta _i)$, where $n_i$ and $\vartheta _i$
are real time-dependent variables.  Let's introduce the population difference $\Delta
=n_0-n_1$ and the phase $\Theta =2\left( {\vartheta _0-\vartheta _1}
\right)+2\left( {\beta _0-\beta _1} \right)t$.  Then the system (5) turns into
equations
\begin{equation} \label{07}
\begin{array} {c}
\dot \Delta =-B{\displaystyle {\left( {1-\Delta ^2} \right)}}\sin\Theta +F\left( t \right)
{\displaystyle {\sqrt {1-\Delta ^2} \over 2}}\sin {\displaystyle{ \Theta  \over 2}},
\\*[9pt]
\dot \Theta =-\Omega +2A\Delta +2B\Delta \cos \Theta -F\left( t \right)
{\displaystyle{\Delta  \over {\sqrt {1-\Delta ^2}}}}\cos {\displaystyle {\Theta  \over 2}}
+G\left( t \right),
\end{array}
\end{equation}
where the following notations have been introduced: $\Omega =2\left( {\beta _1-
\beta _0} \right)+\lambda \hbar ^{-1}\left( {J_{11}-J_{00}} \right)$, $A={{\lambda
\hbar ^{-1}\left( {4J_{01}-J_{00}-J_{11}} \right)} \mathord{\left/ {\vphantom
{{\lambda \hbar ^{-1}\left( {4J_{01}-J_{00}-J_{11}} \right)} 2}} \right. \kern-
\nulldelimiterspace} 2}$, $B=\lambda \hbar ^{-1}J_{01}$, $F\left( t \right)=4\hbar
^{-1}V_{01}\left( t \right)$ and $G\left( t \right)=2\hbar ^{-1}\left( {V_{00}\left( t
\right)-V_{11}\left( t \right)} \right)$.  All these quantities have the same
dimensionality, namely that of the frequency.  If the nonlinearity parameter vanishes,
$\lambda =0$, then Eqs. (7) become equivalent to the well-known Bloch equations
\cite {AE75}.

The nonlinear Bloch equations (7) can be considered as a pair of canonical equations for
the conjugated variables $\Delta ,\Theta $ of the non-autonomous system with one
degree of freedom with the Hamiltonian function $H=H_0+H_1\left( t \right)$, where
the unperturbed Hamiltonian is
\begin{equation} \label{08}
H_0=\Omega \Delta +A\Delta ^2-B{\displaystyle \left( {1-\Delta ^2} \right)}\cos{ \displaystyle
\Theta,}
\end{equation}
and the perturbation is
\begin{equation} \label{09}
H_1\left( t \right)=G\left( t \right)\Delta -F{\displaystyle \left( t \right)\sqrt {1-\Delta ^2}}
\cos{\displaystyle {\Theta  \over 2}}.
\end{equation}
In what follows we refer to the value of the function $H_0$ as the energy of the system.
In the absence of perturbation the system (7) has two trivial stationary solutions,
$\Delta =\pm 1$ and $\Theta $ arbitrary, that correspond to the symmetric
eigenstates $\Phi _0$ and $\Phi _1$ respectively, and two non-trivial fixed points
$\Delta _0=-{\Omega  \mathord{\left/ {\vphantom {\Omega  {2\left( {A+B}
\right)}}} \right. \kern-\nulldelimiterspace} {2\left( {A+B} \right)}}$, $\Theta =0$
or $\Theta =2\pi $, that correspond to the pair of self-trapped states that below will be
called the stationary states.  These states are divided from the bulk of the phase space by
a separatrix (see Fig. 1).

For the future numerical calculations we need to specify the parameters of the
unperturbed Hamiltonian (8).  We have chosen the following set of values: $\Omega
=5.388$, $A=1.902$, and $B=2.022$.  With this choice the stationary states which are
located at $\Delta _0=-0.686$ correspond to the minimal energy of the system $E_-=-
3.871$, the separatrix coincides with the isoenergetic line $E=E_s=-3.486$, and the
maximal energy $E_+=7.290$ corresponds to the line $\Delta =1$.

\section{The harmonic perturbation}

We shall assume that the even (diagonal) perturbation is absent, $G(t)=0$, and the odd
(non-diagonal) has the form
\begin{equation} \label{10}
F\left( t \right)=F\sin \left( {\omega t+\phi } \right).
\end{equation}
The numerical experiments show, that for given values of frequency $\omega$ and
initial phase $\phi$ of the perturbation there is a threshold value $F_c\left( {\omega
,\phi } \right)$ of its amplitude such that for $F<F_c$ the phase trajectory of the
system remains indefinitely within one loop of the separatrix, whereas for $F>F_c$ the
phase trajectory crosses the separatrix, does it repeatedly and may come close to the
opposite stable point.  The dependence of $F_c\left( {\omega ,\phi } \right)$ on the
initial phase is feeble: the relative variations of the thresholds due to the change of the
initial phase have the order of a few percent.  Hence for the time being we ignore this
dependence and shall speak only about the dependence $F_c\left( \omega  \right)$
(see Fig. 2).

The abrupt change of the character of motion at a certain threshold value of the
perturbation magnitude strongly indicates on the onset of the global stochasticity
that comes from the overlap of resonances \cite {Ch79,LL92} and the destruction of the
noble tori \cite {G79,LL92}.  This is indeed the case: by taking some phase point at
the separatrix for the initial conditions, one can see that at the same threshold values
$F_c\left(\omega  \right)$ the stochastic layer around the separatrix explodes and covers
the vicinity of the stable states.  However, even at rather small excesses of $F$ over the
threshold the crossing of separatrix comes fast, after about ten periods of the field.
At these times the chaotic nature of the system's dynamics remains concealed: the motion
seems regular and rather simple.  Therefore we can try to explain the behaviour of
the dependence $F_c\left( \omega  \right)$ in the frame of regular dynamics.

Specific features of the perturbing terms in Eqs. (7) create technical complications
that are irrelevant to the nature of the phenomenon.  The main qualitative features of the
separatrix crossing under the influence of the harmonic field could be explained with a
toy model - the one-dimensional Duffing oscillator with the equation of motion
\begin{equation} \label{11}
\ddot x+x-x^3=F\sin \omega t
\end{equation}
and the initial conditions $x\left( 0 \right)=0, \,\dot x\left( 0 \right)=0$.  This
model also has a stable point, surrounded by a separatrix.

We assume the frequency detuning $\delta =\omega -1$ to be small, $\left| \delta
\right|<<1$.  If the perturbation $F$ is weak, then the non-linearity of the oscillator
could be neglected, at least in the lowest approximation.  Then the solution of Eq. (11)
has the approximate form
\begin{equation} \label{12}
x\left( t \right)\approx -{F \over \delta }\sin {\delta  \over 2}t\cos \,\left[
{\omega t-{\delta  \over 2}t} \right].
\end{equation}
By assumption that that the oscillator could be linearized in all the range $\left| x
\right|\le 1$, from this law of motion we find a crude estimate for the threshold
of the separatrix crossing, namely $F_c=\left| \delta  \right|$.

Now we improve this estimate by taking into account the nonlinearity.  Let's represent
the motion of the oscillator in the form $x\left( t \right)=A\cos \left( {\omega
t+\varphi } \right)$, where $A$ and $\varphi$ are slowly varying functions of time.
The law (12) corresponds to the equations of motion for the slow amplitude and
phase,
\begin{equation} \label{13}
\dot A=-{F \over 2}\cos \varphi ,\qquad \dot \varphi =-{\delta  \over 2},
\end{equation}
with the initial conditions $A\left( 0 \right)=0$, $\varphi \left( 0 \right)=\pi $.
Let's now replace the eigenfrequency of the oscillator $\Omega _0=1$ in the RHS of
the second of Eqs. (13) by the eigenfrequency of the nonlinear oscillator $\Omega
\left( A \right)$ that depends on the amplitude.  For the model (11) for small $A$ we
have $\Omega \left( A \right)=1-{{3A^2} \mathord{\left/ {\vphantom {{3A^2}
{16}}} \right. \kern-\nulldelimiterspace} {16}}$.  Consequently the evolution of the
system could be described by the system of equations
\begin{equation} \label{14}
\dot A=-{F \over 2}\cos \varphi ,\qquad \dot \varphi =-{\delta  \over 2}-{3 \over
{32}}A^2,
\end{equation}
with the same initial conditions.  The threshold of the separatrix crossing could be found
from the condition that oscillations may reach the saddle points: $\max A(t)=1$.  From
the second of Eqs. (14) it is seen, that if $\delta >0,$ then the phase shift
decreases monotonously, thus decreasing the rate of the amplitude growth.  If $\delta
<{{-3} \mathord{\left/ {\vphantom {{-3} {16}}} \right. \kern-\nulldelimiterspace}
{16}}=-0.188$, then the phase shift increases monotonously while the amplitude
stays below its critical value, $A<1$; and again the rate of the amplitude growth
decreases with time.  But in the band ${{-3} \mathord{\left/ {\vphantom {{-3} {16}}} \right.
\kern-\nulldelimiterspace} {16}}=-0.188<\delta <0$ the phase shift at first grows, then
reaches the maximum and starts to decrease, passing the zero value at some later time
$t_0$.  Consequently, there are two moments in which the amplitude growth rate is
maximal, $t=0$ and $t=t_0$.  Thus one may expect that the dependence $F_c\left(
\omega  \right)$ will have the minimum somewhere in the range ${{13}
\mathord{\left/ {\vphantom {{13} {16}}} \right. \kern-\nulldelimiterspace}
{16}}=0.812<\omega <1$.  The numerical solution of Eqs. (14) shows that this is
true: the minimum of $F_c\left( \omega  \right)$ is reached for $\omega
=0.87$, about the middle of this band (see Fig. 3).

To find the condition for the separatrix crossing, the equations (14) should be solved on
the finite interval of time, while the phase shift reaches the value $\pm {\pi
\mathord{\left/ {\vphantom {\pi  2}} \right. \kern-\nulldelimiterspace} 2}$.  This
could be done in different ways, that may produce analytical estimates for the threshold
values.  We restrict ourselves by the only example.  In the exact resonance (at $\delta
=0$) in the zeroth approximation the amplitude dependence on time is linear,
$A_0={{Ft} \mathord{\left/ {\vphantom {{Ft} 2}} \right. \kern-\nulldelimiterspace}
2}$.  By substitution of this expression in the RHS of the second of Eqs. (14), we
have in the first approximation
\begin{equation} \label{15}
\varphi _1(t)=\pi -{1 \over {128}}F^2t^3.
\end{equation}
The time $t_m$ when the amplitude $A(t)$ reaches the maximum could be found from
the condition $\varphi _1\left( {t_m} \right)={\pi  \mathord{\left/ {\vphantom
{\pi  2}} \right. \kern-\nulldelimiterspace} 2}$.  Hence from the first of Eqs. (14)
in the first approximation we have the threshold value of the perturbation in the exact
resonance:
\begin{equation} \label{16}
F_c(1)={{27} \over {16}}\left[ {\int\limits_0^{{\pi  \mathord{\left/ {\vphantom
{\pi  2}} \right. \kern-\nulldelimiterspace} 2}} {\theta ^{-{2 \mathord{\left/
{\vphantom {2 3}} \right. \kern-\nulldelimiterspace} 3}}\,\cos \theta \,d\theta }}
\right]^{-3}=0.0666
\end{equation}
It agrees with the result of the numerical solution of the system (14) with the accuracy
about 6\%, but differs from the value obtained in the numerical experiment by a factor
about 1.5.

The studied model (11) shares with the system (7) the "soft" character of the
nonlinearity of oscillations around the stable points: the eigenfrequency of oscillations
decrease with the growth of their amplitude.  This common feature is responsible for the
similar behaviour of $F_c(\omega)$ in two models (compare Figs. 2 and 3) -
namely, the presence of non-zero minimum at a frequency somewhat lower than that of
the small oscillations, $\Omega _0.$

Now we return to the case of the chaotic motion of the system above the threshold.  For
the system with the Hamiltonian $H=H_0\left( {\Delta ,\Theta } \right)+V\left(
{\Delta ,\Theta } \right)\sin \omega t$ with a small perturbation $V$ the energy
half-width of the stochastic layer is given by the Melnikov - Arnold integral
\begin{equation} \label{17}
\Delta E=\int\limits_{-\infty }^\infty  {\left( {{{\partial V} \over {\partial \Delta
}}\dot \Delta +{{\partial V} \over {\partial \Theta }}\dot \Theta } \right)}\sin
\omega t\, dt,
\end{equation}
where $\Delta \left( t \right)$ and $\Theta \left( t \right)$ are taken for the
unperturbed motion on the separatrix \cite {Ch79}.  Thus we can expect that the motion
in the phase space will persist inside the domain limited by the isoenergetic line
$E=E_s+\Delta E$.  Since for the Hamiltonian systems the phase volume is conserved,
we may expect the invariant density in the phase space to be uniform. This leads to the
invariant distribution of the energy values
$w\left( E \right)$ of the form
\begin{equation} \label{18}
\begin{array} {c}
w \left( E \right)=\eta {\displaystyle {2 \over {\Omega \left( E \right)}}},
\qquad (E_{-}<E<E_s),\\*[9pt]
w \left( E \right)=\eta {\displaystyle {1 \over {\Omega \left( E \right)}}},
\qquad (E_s <E<E_s+\Delta E),
\end{array}
\end{equation}
where $\eta$ is the normalization constant, and the factor "2" in the first line accounts
for the double degeneracy of the energy states.  The comparison of this distribution with
the one obtained in the numerical experiments is shown in Fig. 4.  The general
agreement is clearly present, in spite of rather large value of the perturbation magnitude.
The discrepancy between the distributions for the energy values around and above
$E_s+\Delta E$ is due to the borderline resonances of the stochastic layer and
could be anticipated.

If we define the vicinity of the stable state by the condition $E<E_*$, then the average
fraction of time spent in this domain is
\begin{equation} \label{19}
\mu \left( {E_*} \right)=\eta { \int \limits _{E_-}^{E_*}} {{{dE} \over {\Omega
\left( E \right)}}}\approx \eta {{(E_*-E_{-})} \over {\Omega _0}}
\end{equation}
and the average transition time from vicinity of one of the self-trapped states to vicinity
of the other is about
\begin{equation} \label{20}
T \sim {\tau  \over {\mu \left( E_{*} \right)}}
\end{equation}
where $\tau$ is the energy relaxation time.  For the over-threshold perturbation amplitude the
latter has value about the period of the harmonic field.

\section{The broadband perturbation}

The threshold character of the separatrix crossing in the harmonic field stems from the
termination of the resonant energy absorption before the value $E_s$ is achieved.  If the
perturbation is a broadband noise, then the energy absorption can go on indefinitely at
arbitrarily small field amplitude.  The problem of the energy absorption from the external
noise has been studied extensively as a part of the theory of the escape over the potential
barriers in dissipative systems in contact with the heat bath.  The averaged evolution of
the system can be described as a process of diffusion on the energy axis.  The equation
that governs this process could be derived from the Fokker - Planck equation
\cite {IM86,LG91}.  It is inconvenient, however, to adjust the known results to our
case since our Hamiltonian (7,8) has rather unusual structure.  Instead we derive the
equation for the energy diffusion considering the system {\it formally} as quantum one,
starting from quantum kinetic equations and going to the classical limit $\hbar \to 0$
in the result (cf. \cite {LO86,ESh96}).

Let's consider a quantum system with the unperturbed Hamiltonian $\hat H_0$ with
one degree of freedom and discrete energy spectrum under the perturbation
$\hat V\xi \left( t \right)$ where $\hat V$ depends on the dynamical variables of
the system and $\xi \left( t \right)$ is a stationary weak broadband noise specified by
its spectral density $S\left( \omega  \right)$.  The state of the system could be
described by the probabilities $\rho _n$ of finding it in the quantum state $\left| n
\right\rangle $.  The evolution of these probabilities obeys the system of master equations
\begin{equation} \label{21}
{{d\rho _n} \over {dt}}=-\rho _n\sum\limits_{k=-n}^\infty  {\dot
W_{n,n+k}}+\sum\limits_{k=-n}^\infty  {\rho _{n+k}\dot W_{n+k,n}}.
\end{equation}
The rates of transitions $\dot W_{n+k,n}$ are determined by the perturbation theory formula
\begin{equation} \label{22}
\dot W_{n,n+k}={{2\pi } \over {\hbar ^2}}\left| {V_{n,n+k}} \right|^2S\left( {-
\omega _{n,n+k}} \right),
\end{equation}
where $V_{n,n+k}$ are the matrix elements of the perturbation and $S\left( {-\omega
_{n,n+k}} \right)$ is the spectral density of noise at the frequency of transition.  Let's
take the probabilities to be functions not on the level number $n$, but on its energy:
$\rho _n\to \rho \left( {E_n} \right)$.  In the quasiclassical case the energy
spectrum of the system could be related to the frequency of its classical motion at a
given energy $\Omega \left( E \right)$:
\begin{equation} \label{23}
E_{n+k}=E_n\pm \hbar \Omega \left( {E_n\pm k{{\hbar \Omega } \over 2}}
\right),
\end{equation}
and the matrix elements of the perturbation $V_{n,n+k}$ could be replaced by the
Fourier components of the unperturbed motion of the dynamical variable that corresponds
to the operator $\hat V$: if
\begin{equation} \label{24}
V\left( t \right)=\sum\limits_{k=-\infty }^\infty  {V_ke^{-ik\Omega t}},
\end{equation}
then
\begin{equation} \label{25}
V_{n,n+k}\to V_k\left( {E_n+k{{\hbar \Omega } \over 2}} \right).
\end{equation}
We assume $\rho \left( E \right)$ to be a smooth function, and expand its value to
the terms of the second order in $\hbar $ and the values of $E_{n+k}$ and
$V_{n,n+k}$ to the first order in $\hbar $.  After the substitution of these expansions
in Eq. (21) and going to the limit $\hbar \to 0$ we obtain purely classical
equation.  At this point we restrict our consideration to the case of the white noise with
the constant spectral density $S\left( \omega  \right)=1$.  For this case we have
\begin{equation} \label{26}
{{\partial \rho } \over {\partial t}}=\Omega {\partial  \over {\partial E}}\left[
{D\left( E \right){{\partial \rho } \over {\partial E}}} \right],
\end{equation}
where the energy diffusion coefficient $D\left( E \right)$ is given by the expression
\begin{equation} \label{27}
D\left( E \right)=2\pi \Omega \left( E \right)\sum\limits_{k=1}^\infty
{k^2V_k^2\left( E \right)}={1 \over 2}\int\limits_0^{{{2\pi } \mathord{\left/
{\vphantom {{2\pi } \Omega }} \right. \kern-\nulldelimiterspace} \Omega }} {\left(
{{{dV} \over {dt}}} \right)^2dt}.
\end{equation}
The energy dependence of the diffusion coefficient $D$ for the unperturbed system (7)
with the perturbation $V(\Delta, \Theta)= \sqrt {1-\Delta ^2}\cos (\Theta / 2)$ is
shown in Fig. 5.  We note a discontinuity of $D(E)$ at the separatrix value of
energy: $D(E_s +0)=2D(E_s -0)$.  This jump is not a direct consequence of the
presence of the separatrix, but reflects both the global structure of the phase space and
the behaviour of the perturbation $V$ in the neighbourhood of the separatrix.

Since the classical distribution in energy $w\left( E \right)$ is connected to the
probability density $\rho \left( E \right)$ by the relation $w\left( E \right)=\rho
 \left( E \right)\left[ {\hbar \Omega \left( E \right)} \right]^{-1}$, we have the
equation for the $w\left( E \right)$ in the form
\begin{equation} \label{28}
{{\partial w} \over {\partial t}}={\partial  \over {\partial E}}\left( {D\left( E
\right){\partial  \over {\partial E}}\left( {w\Omega } \right)} \right).
\end{equation}

The stationary solution of Eq. (28) has the form given by Eq. (18), but the second line
holds now in all the range $E_s < E < E_{+}$.  If the system initially was in one of the
stable states, then in course of the energy diffusion it has an opportunity to migrate to
vicinity of the contrary stable state.  The characteristic time of the average transfer
is determined by the relaxation time of the distribution to its stationary value; from
Eq. (26) it can be estimated as
\begin{equation} \label{29}
T \sim {{\left( {E_+-E_-} \right)^2} \over
{\left\langle {\Omega D} \right\rangle }},
\end{equation}
where the angular brackets denote averaging over the energy.

\section{Conclusions}

The main question addressed by this paper is: "If the system, that is described by the
one-dimensional non-linear Scr\"odinger equation with the potential of the symmetric
double well, is initially in one of the lowest asymmetric (self-trapped) states, then can the
time-dependent perturbation transfer the system completely to the opposite asymmetric
mode?"  The answer to it comes to be "Yes, but almost completely and only by chance".

For the harmonic perturbation with the amplitude $F$ that exceeds the threshold
$F_c (\omega)$, the system's motion is captured in the stochastic layer that embraces
both domains of libration around the stable states and a strip around the separatrix.
When moving in this domain, the system can come arbitrarily close to the opposite
asymmetric state.  However, the nature of this process is purely chaotic and, thence,
unpredictable.  There is no way to create for the nonlinear system the "$\pi$ - pulse"
\cite{AE75} that will transfer the system unambiguously from one of the stable states
to another.  Lastly we note, that the threshold magnitude of the perturbation is only
numerically small in comparison with the depth of the self-trapping wells: to make the
transfer possible, the system must be perturbed strongly.

For the system under the influence of the white noise (or, generally speaking, any
sufficiently broadband noise) the process of the energy diffusion eventually spreads
the probability density over all phase space of the system $H_0$.  In this case the
system can occasionally come close to the opposite stable state.  We note, however, that
the probability of finding the system within one of the self-trapping wells is rather
small; with our standard set of parameters it is only about few percent.

The main approximation of our calculations consisted in the truncation of the
expansion of the wave function to just two modes (see Eq. (4)).  It was justified
by the high quality of this approximation in representation the unperturbed
self-trapped states \cite {OK+99}.  Whether this accuracy will hold for the seriously
perturbed system is quite a different question.  For the harmonic perturbation of
moderate amplitude the system stays locked within the narrow energy domain (see Fig. 4),
and the addition of contributions of modes $\Phi_i$ with $i \ge 2$, that will
lead to the extension of the energy space of the system, will have little influence.

The situation may be different for the broadband noise, where the system can
reach any point in the phase space.  However the main contribution to the energy
diffusion coefficient comes from the frequencies that are lower than $\Omega_0$:
in particular, at the separatrix they contribute about 0.66 of the total value.
Thus if the spectrum of noise has cutoff of high frequencies just above
the $\Omega_0$, then the energy diffusion ceases at the energy $E_{h}>E_s$ at
which $\Omega (E_h)=\Omega_0$ (for our set of parameters $E_h=-0.356$)
and the system stays locked within the restricted energy domain.  Then
in parallel to the case of the harmonic field we can conclude on the
unimportance of the extension of the energy space by addition of higher modes.

\section*{Acknowledgment}

The authors are grateful to Prof. R. Sammut and Dr. A.V. Buryak for the introduction in
the field and to Prof. Yu.M. Romanovsky for specifying the studied problem.  The
permanent informational support of this work by Dr. E.A. Ostrovskaya and Dr. D.G.
Luchinsky and discussions with them were invaluable.

This work was financially supported by the Education and Science Center "Fundamental
Optics and Spectroscopy" (in the frame of the program "Integration") and by the Russian
Federal grant \# 96-15-96476 for the support of outstanding scientific schools.

\newpage

\section*{Figures captions}

\noindent FIG.1.  Phase trajectories of the system (7) in the absence of perturbations
on the plane $\Delta -\Theta $ for different values of energy: A - $E=-3.7$,
B - $E=E_s=-3.487$, C - $E=0$, D - $E=5$.

\vspace{3mm}

\noindent FIG.2.  Dependence of threshold amplitudes $F_c$ of the non-diagonal
harmonic perturbation (10) on the relative frequency ${\omega} / {\Omega_0}$
for the nonlinear Bloch equations (7) with the initial phase $\phi =0$.

\vspace{3mm}

\noindent FIG.3.  Dependence of threshold amplitudes $F_c$  of the perturbation on the
frequency $\omega$ for the Duffing oscillator (11). The dashed line - the zeroth
(linear) approximation, $\circ$ - estimates found from Eqs. (14) (solid
line is an eyeguide); $\bullet$ - results of the numerical experiment.

\vspace{3mm}

\noindent FIG.4.  Distribution $w$ of system's values of energy under the influence of
the over-threshold harmonic perturbation with $\omega = \Omega_0 = 2.887$ and
$F=2F_c=0.2$.  Dashed line -theoretical distribution Eq. (18), solid
line - results of the numerical experiment.

\vspace{3mm}

\noindent FIG.5. Dependence of the coefficient of the energy diffusion $D$ under
the influence of the white noise of the unit spectral density on the energy
$E$.

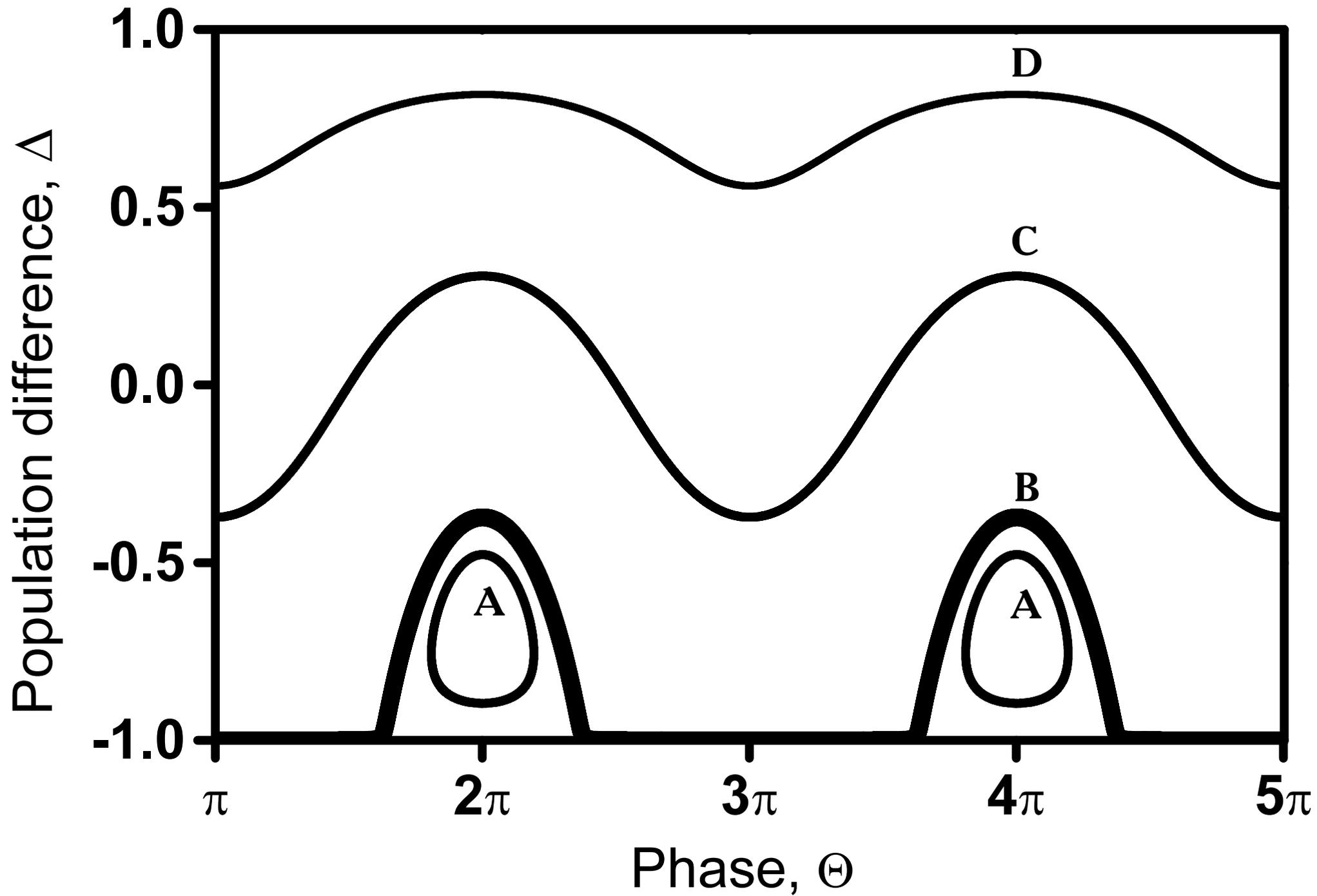
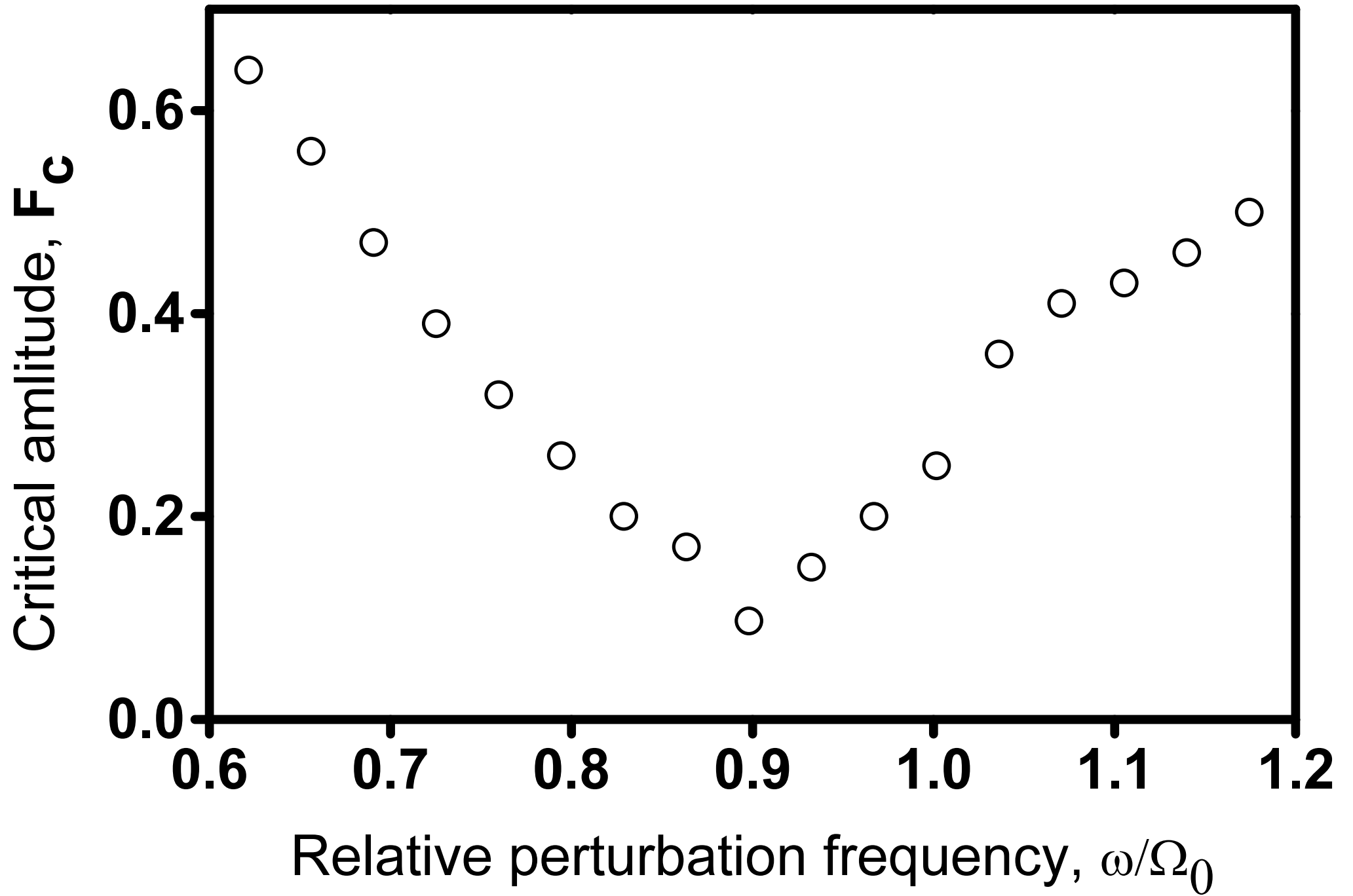
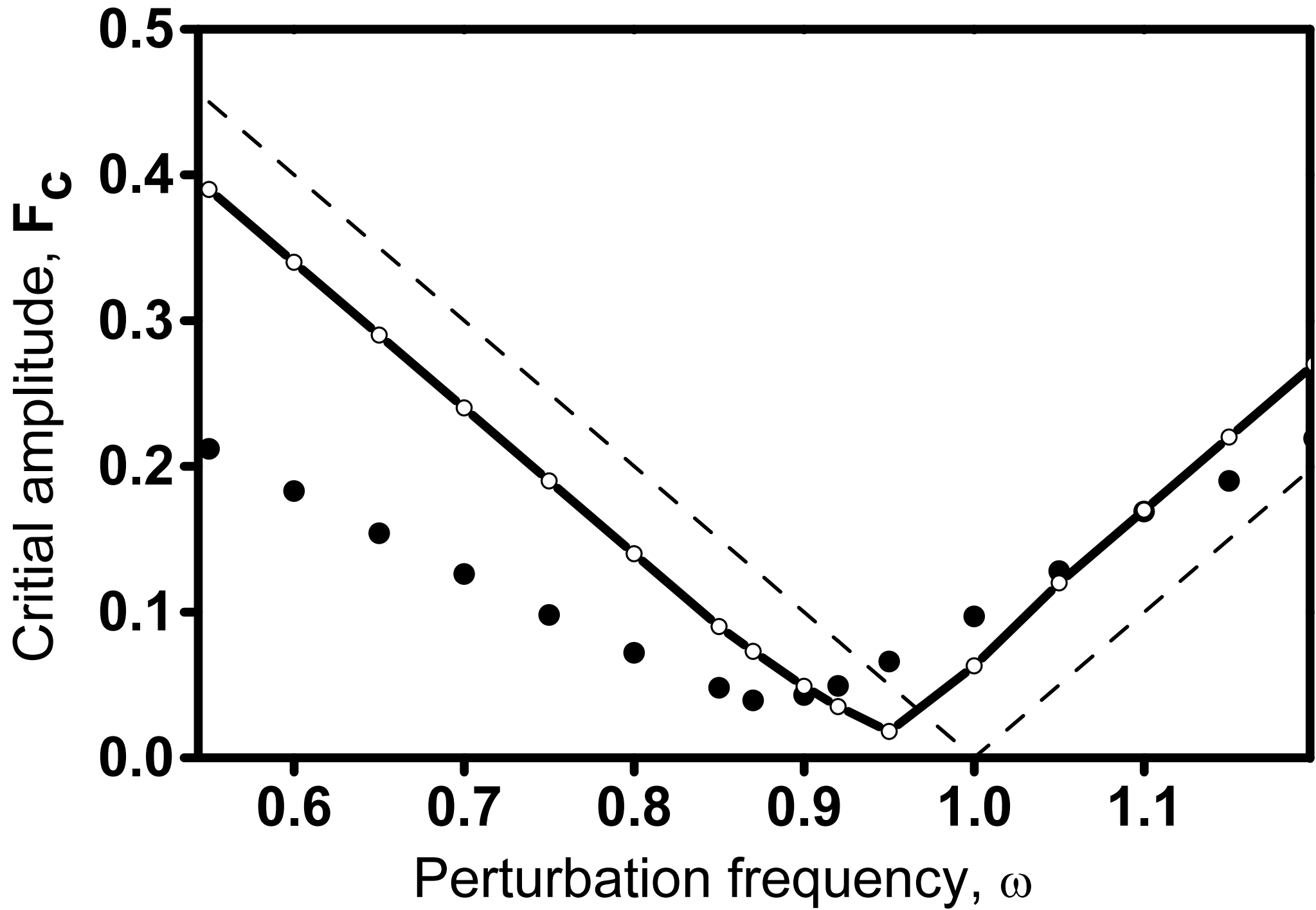
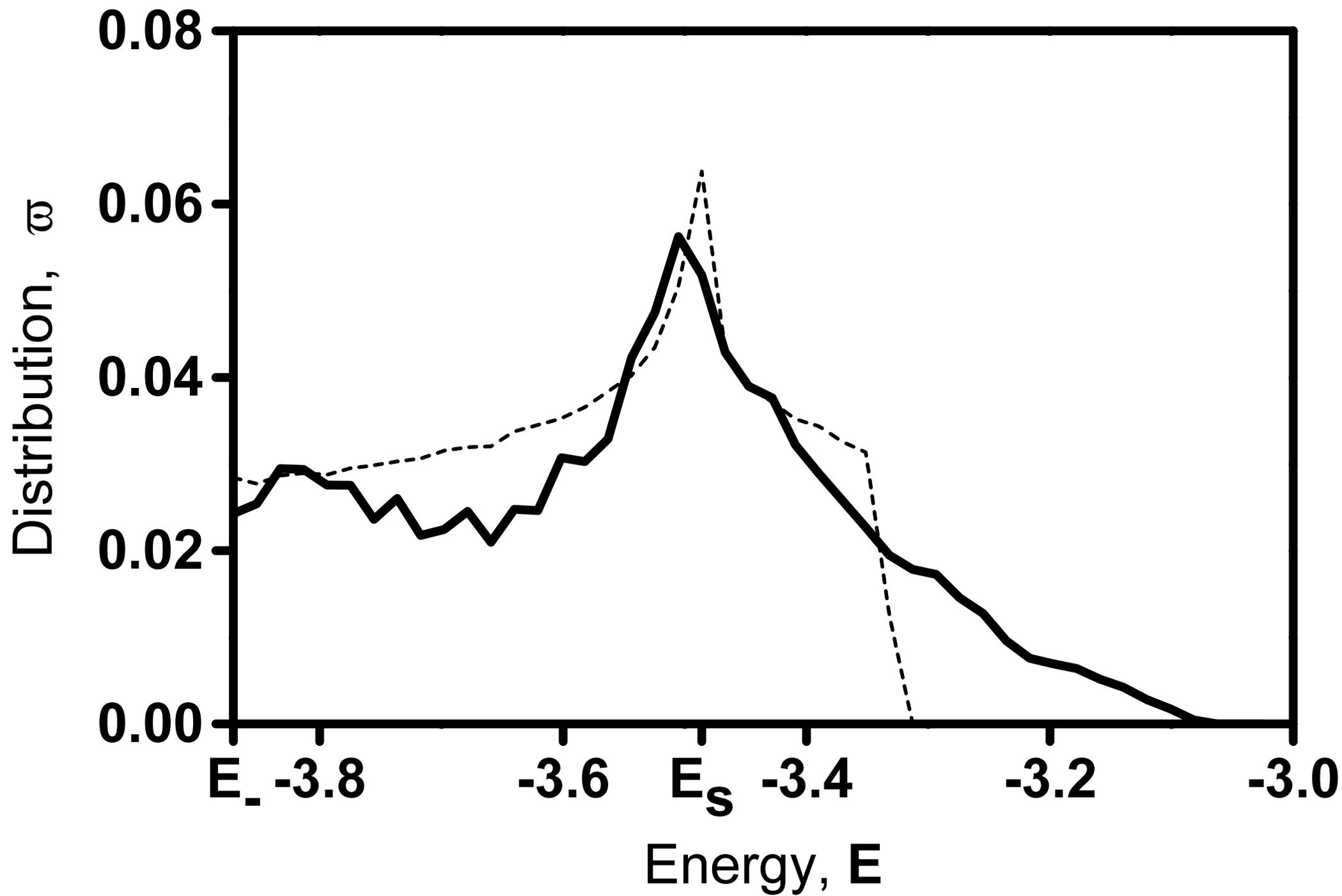
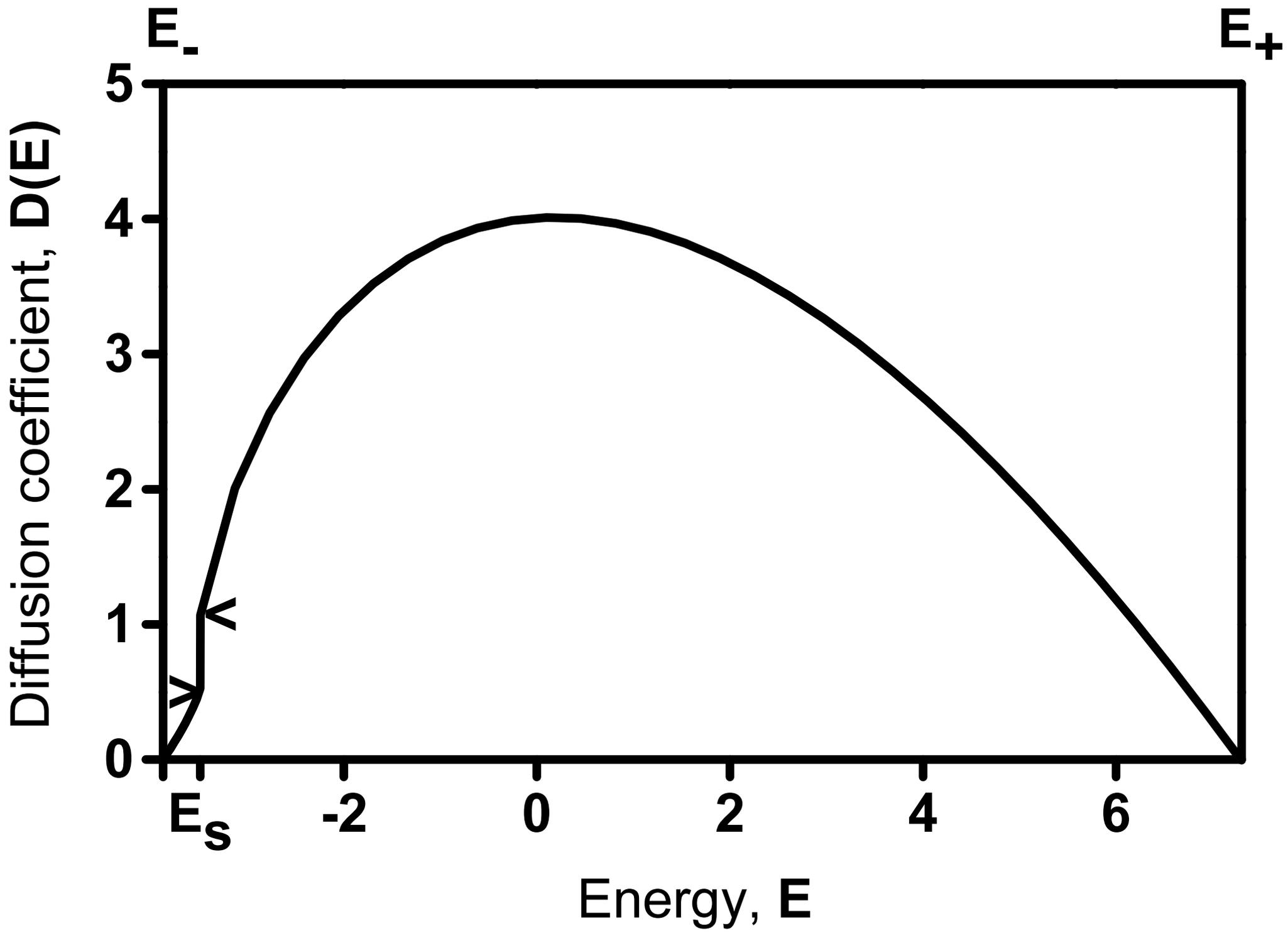
\end{document}